\documentclass[10pt]{article}

\title{Static wormholes on the brane inspired by Kaluza-Klein gravity }
\author{J. Ponce de Leon\thanks{E-Mail:
jpdel@ltp.upr.clu.edu, jpdel1@hotmail.com}  \\Laboratory of Theoretical Physics, 
Department of Physics\\ 
University of Puerto Rico, P.O. Box 23343,  
San Juan,\\ PR 00931, USA}
\date{October 2009}

\usepackage{graphicx}
\graphicspath{%
    {converted_graphics/}
    {F:/WormholesJCAP/converted_graphics/}
}
\begin{document}

\maketitle
\begin{abstract}
We use static solutions of $5$-dimensional Kaluza-Klein gravity to generate several classes of static, spherically symmetric  spacetimes which are analytic solutions to the equation $^{(4)}R = 0$, where $^{(4)}R$ is the four-dimensional Ricci scalar. 
In the  Randall $\&$ Sundrum  scenario they can be interpreted as vacuum solutions on the brane.  The solutions  contain the Schwarzschild black hole, and generate new families of traversable Lorenzian wormholes as well as nakedly singular spacetimes. They generalize a number of previously known solutions in the literature, e.g.,  the temporal and spatial Schwarzschild solutions of braneworld theory as well as the class of self-dual Lorenzian wormholes.
A major departure of our 
solutions  from Lorenzian wormholes {\it  a la} Morris and Thorne is that, for certain values of the parameters of the solutions, they contain three spherical surfaces (instead of one) which are extremal and have finite area. Two of them have the same size, meet the  ``flare-out" requirements, and show the typical violation of the energy conditions  that characterizes a  wormhole throat. The other extremal sphere is ``flaring-in" in the sense that its sectional area is a local maximum and the weak, null and dominant energy conditions are satisfied in its neighborhood. After bouncing back  at this second surface a traveler  crosses into another space which is the double of the one she/he started in. Another interesting feature is that  the size of the throat can be less than the Schwarzschild radius $2 M$, which no longer defines the horizon, i.e., to a distant observer a particle or light falling down crosses the Schwarzschild radius in a finite time.

\end{abstract}

\medskip



\newpage
\section{Introduction}

The idea that remote parts of the universe, or even that {\it different}
universes, could be connected by smooth bridges or tunnels  constitutes a very intriguing concept, which for many decades has captivated  the imagination of science fiction lovers \cite{WHwiki}, \cite{TTwiki}. In physics, spacetimes containing  such bridges (called ``wormholes" by J.A. Wheeler) appear in general relativity as  solutions of the Einstein field equations.  Nearly half a century ago,  the concept of wormholes 
lead Wheeler to the discussion of  topological entities called geons\footnote{In \cite{geons} Wheeler provides the first diagram of a wormhole as a tunnel connecting two openings in different regions of spacetime.} \cite{geons} and to the conception of geometrodynamics \cite{Geometrodynamics}, where ``matter comes from no matter", and ``charge comes from no charge".

Lately, after the fundamental papers by  Morris, Thorne and Yurtsewer \cite{Morris 1} and Morris and Thorne \cite{Morris 2}, the notion of traversable Lorentzian wormholes has gained much attention within the physics community. These authors  showed that such wormholes  could, in principle, allow humans not only to travel  between universes, or distant parts of a single
universe, but also to construct time machines. It has been suggested that black holes and wormholes
are interconvertible. In particular that stationary
wormholes could be 
possible final states of black-hole evaporation \cite{Hayward}. Also, that astrophysical accretion of ordinary matter could convert wormholes into black holes \cite{Novikov} (this issue has recently been discussed in literature and different approaches give, in general,  different results - see \cite{Kuhfittig1}, \cite{Sergey}).

Today, it is well known that a wormhole geometry can only appear as a solution of the Einstein field equations if the energy-momentum tensor (EMT) of the matter supporting such a geometry violates the null energy condition (NEC) at least in the neighborhood of the wormhole throat \cite{Visser 1}-\cite{Visser 3} (matter that violates the NEC is usually called {\it exotic}).  Although in general relativity there are many examples of matter that are consistent with wormhole spacetimes (see, e.g.,  \cite{Barcelo}-\cite{Kuhfittig2}), none of them are observable in the real world of astrophysics\footnote{To be fair, we should mention that the solutions discussed in \cite{Barcelo}-\cite{Sushkov 1} were obtained as early as in  1973 by Bronnikov \cite{AP} and also by Ellis \cite{Ellis} describing wormhole solutions with a massless, minimally coupled phantom scalar field. }: ``all these spacetimes still remain in the domain of fiction" \cite{Dadhich}. This is 
a very important and challenging issue in wormhole physics, which is known as the  ``exotic matter problem". There are numerous attempts  at solving this issue in the literature.  Some consider alternative theories of gravity \cite{Nandy}-\cite{Hochberg1} or invoke quantum effects in curved spacetimes, considering wormholes as semiclasical objects \cite{Hochberg2}, \cite{Remo}.  Recently, a general no-go theorem was proved by Bronnikov and Starobinsky showing the absence of wormholes in scalar-tensor gravity without ghosts \cite{BS}. 

A different approach to this issue arises  in the context of higher-dimensional theories. In Kaluza-Klein gravity the exotic matter necessary for the formation of a wormhole can appear from the off-diagonal elements of the metric (the gauge fields) and from the $\gamma_{5 5}$ component of the metric (the scalar field), rather than coming from some externally given exotic matter \cite{Singleton}. Also the so-called 
Einstein-Gauss-Bonnet theory admits wormhole solutions that would not violate the NEC
provided the Gauss-Bonnet coupling constant is negative \cite{Bhawal}, \cite{Dotti}. In addition, it has been proposed that braneworld gravity provides a natural scenario for the existence of traversable wormholes \cite{Bronnikov 1}, \cite{Lobo}. This is because the local high-energy  effects and non-local corrections from the Weyl curvature in the bulk could lead to an effective violation of the NEC on the brane even when the standard-model fields satisfy the energy conditions.
In this paper we adhere to the latter framework and develop a number of spherically symmetric, static Lorentzian wormholes which are analytic solutions to the equations on the brane. 

In the  Randall $\&$ Sundrum  braneworld scenario \cite{Randall2} the effective equations for gravity in $4D$ were derived by Shiromizu, Maeda and Sasaki \cite{Shiromizu}.  In vacuum, when matter on the brane is absent and the $4$-dimensional cosmological constant vanishes, these equations reduce to\footnote{Throughout the paper we use the conventions and definitions of Landau and Lifshitz \cite{Landau} and set  $G = c = 1$.} 
\begin{equation}
\label{SMS equations}
^{(4)}G_{\alpha \beta} =  - \epsilon E_{\alpha\beta},
\end{equation}
where $^{(4)}G_{\alpha \beta}$ is the usual  Einstein tensor in $4D$; 
$\epsilon$ is taken to be $- 1$ or $+ 1$ depending on whether the extra dimension is spacelike or timelike, respectively;  
$E_{\alpha \beta}$ is the projection onto the brane of the Weyl tensor in $5D$. Explicitly,  $E_{\alpha\beta} = {^{(5)}C}_{\alpha A \beta B}n^An^B$, where $n^{A}$ is the $5D$ unit vector $(n_{A}n^{A} = \epsilon)$ orthogonal to the brane. This quantity connects the physics in $4D$ with the geometry of the bulk.

The vacuum field equations on the brane (\ref{SMS equations}) are formally equivalent to the Einstein equations of general relativity with an effective EMT given by  $T_{\alpha \beta} = - \left(\epsilon/8 \pi\right) E_{\alpha \beta}$. The crucial point here is that due to its geometrical nature, $E_{\alpha \beta}$ does not have to satisfy the energy conditions applicable to ordinary matter. In fact, there are a number of examples in the literature where the effective EMT  corresponds to exotic matter on the brane \cite{Vollick}, \cite{Bronnikov 2}. Thus, $E_{\alpha\beta}$ is the most natural ``matter" supporting wormholes \cite{Bronnikov 1}.  

However, the set of equations (\ref{SMS equations}) does not form a closed system in $4D$, because $E_{\alpha \beta}$ is unknown without specifying, {\it both} the metric in $5D$, and the way the $4D$ spacetime is identified, i.e., $n^{A}$.  The only truly general thing  we know is that $E_{\alpha\beta}$ is traceless. Therefore, the only quantity that can be unambiguously specified on the brane is the trace of the curvature scalar  $^{(4)}R = {^{(4)}R}^{\alpha}_{\alpha}$.  In particular, in empty space 
\begin{equation}
\label{field eqs. for empty space}
^{(4)}R = 0.
\end{equation}
In this paper we investigate spherically symmetric solutions to this equation of the form
\begin{equation}
\label{the metric under study}
ds^2 = A^2(r) dt^2 - B^2(r) dr^2 - r^2 C^2(r) \; d\Omega^2, 
\end{equation}
where $d\Omega^2 = d\theta^2 + \sin^2{\theta} d\phi^2$ is the line element on a unit sphere. In these coordinates the field equation (\ref{field eqs. for empty space}) can be written as (a prime denotes differentiation with respect to $r$)
\begin{equation}
\label{field equation for the general spherical metric}
\frac{A''}{A} + 2\left(\frac{A'}{A} - \frac{B'}{B}\right)\left(\frac{1}{r} + \frac{C'}{C}\right) - \frac{A' B'}{A B} + \frac{1}{r^2}\left(1  - \frac{B^2}{C^2}\right) + 2 \frac{C''}{C} + \frac{C'}{C}\left(\frac{6}{r} + \frac{C'}{C}\right) = 0.
\end{equation}
In curvature coordinates, i.e., in coordinates where $C(r) = 1$, this is a second-order differential equation for $A(r)$       and a first-order one for $B(r)$. Therefore, it has a nondenumerable infinity of solutions parameterized by some arbitrary function of the radial coordinate $r$ \cite{Visser}. 

In curvature coordinates the simplest (non-trivial) solutions to (\ref{field equation for the general spherical metric}) are obtained by 
setting either $A = 1$ or $B = 1$. The former case yields the spatial-Schwarzschild wormhole \cite{Dadhich}, while the latter one gives $A^2 = \left(1 - m/r\right)^2$, which is a black hole with total mass $M = m$ and a horizon  at $r = M$. The next simple solution is the Schwarzschild metric
\begin{equation}
\label{Schwarzschild metric}
ds^2 = \left(1 - \frac{2 m}{r}\right) d t^2 - \left(1 - \frac{2 m}{r}\right)^{- 1} d r^2 - r^2 d\Omega^2. 
\end{equation}
If one chooses either $A(r)$ or $B(r)$ as in the Schwarzschild metric, then the  asymptotic flatness of the solutions is guaranteed, and the vacuum braneworld solutions contain the Schwarzschild spacetime as a particular case. This choice generates  the ``temporal Schwarzschild" metric \cite{Germani}, \cite{CFM},  
\begin{equation}
\label{T-Schw exterior}
ds^2 = \left(1 - \frac{2{{m}}}{r}\right) dt^2 - \frac{(1 - 3{{m}}/2r)}{(1 - 2{{m}}/r)[1 - (3{{m}}/2r)\;\sigma]}dr^2 - r^2 d\Omega^2,
\end{equation}
and the ``spatial Schwarzschild"  metric \cite{CFM}
\begin{equation}
\label{S-Schw exterior}
ds^2 = \frac{1}{\alpha^2}\left(\alpha - 1 + \sqrt{1 - \frac{2 \alpha m}{r}}\right)^2\;dt^2 - \left( 1 - \frac{2 \alpha m}{r}\right)^{- 1}dr^2 - r^2 d\Omega^2,
\end{equation}
where $\sigma$ and  $\alpha$  are dimensionless constant parameters. For $\sigma = 1$, $\alpha = 1$ the corresponding solutions reduce to the Schwarzschild spacetime. The above metrics  have thoroughly been discussed in different contexts:  as braneworld black holes  \cite{Bronnikov 1}, \cite{CFM}; as possible non-Schwarzshild exteriors for spherical stars on the brane \cite{Germani}, \cite{JPdeL 1}, \cite{JPdeL 2}, and as wormhole spacetimes \cite{Bronnikov 2}.

In this work we construct several families of new solutions to $^{(4)}R = 0$ of the form (\ref{the metric under study}), with $C(r) \neq 1$.  The new solutions are specifically designed so that they contain the Schwarzschild black hole and generalize the temporal and spatial Schwarzschild solutions mentioned above.  They  generate new families of traversable Lorenzian wormholes as well as nakedly singular spacetimes. The solutions are obtained by demanding that the spacetime must contain, instead of the Schwarzschild  geometry as in (\ref{T-Schw exterior})-(\ref{S-Schw exterior}), a simple static solution to $^{(4)}R = 0$  that follows from  $5$-dimensional Kaluza-Klein gravity (see Eq. (\ref{Kramer-like solution}) bellow).

Some interesting features  of our models are that, for certain values of the parameters of the solutions, (i) the size of the throat can be less than the Schwarzschild radius $2 M$, which no longer defines the horizon, i.e., to a distant observer a particle or light falling down crosses the Schwarzschild radius in a finite time; (ii) they contain three spherical surfaces (instead of one as in Lorentzian wormholes {\it a la } Morris and Thorne) which are extremal and have finite area. Two of them have the same size and meet the  ``flare-out" requirements\footnote{The flare-out condition defines the throat of a wormhole as a closed two-dimensional spatial minimal hypersurface, i.e., as an extremal surface of minimal area \cite{Visser 3}. Thus, as seen from outside, a wormhole
entrance is a local object like a star or a black hole \cite{Lemos}.}, and show the typical violation of the energy conditions  that characterizes a  wormhole throat. The other extremal sphere is ``flaring-in" in the sense that its sectional area is a local maximum and the weak, null and dominant energy conditions are satisfied in its neighborhood. After bouncing back  at this second surface a traveler  crosses into another space which is the double of the one she/he started in.

The paper is organized as follows. In section $2$ we present our solutions on the brane and discuss their physical interpretation. In section $3$ we construct symmetric traversable wormholes from configurations which only have  one asymptotic region. In section $4$ we give a brief summary of our results.  

\section{Schwarzschild-like solutions on the brane from Kaluza-Klein gravity}

In this  section we construct several new families of analytic solutions to  the brane field equation $^{(4)}R = 0$, of the form (\ref{the metric under study}),  which are inspired by  five-dimensional  Kaluza-Klein gravity and generalize the Schwarzschild-like spacetimes (\ref{T-Schw exterior}) and (\ref{S-Schw exterior}). 

In Kaluza-Klein gravity there is only one family of spherically symmetric exact solutions to the field equations $R_{A B} = 0$ which are asymptotically flat, static and independent of the ``extra" coordinates (see, e.g., Ref. \cite{JPdeL3} and references therein). In five dimensions, in the form given by Kramer \cite{Kramer},  they are described by the line element
\begin{equation}
\label{Kramer's solution in 5D}
dS^2 = f^a \; dt^2 - f^{- (a + b)}\; dr^2 - r^2 f^{(1 - a - b)}d\Omega^2 - f^b dy^2,
\end{equation}
where $y$ is the coordinate along the fifth dimension; 
\begin{equation}
\label{definition of f}
f = 1 - \frac{2 m}{r},\;\;\;m = \mbox{constant},
\end{equation}
and $a$, $b$ are parameters satisfying the consistency relation
\begin{equation}
\label{condition on a and b}
a^2 + ab + b^2 = 1.
\end{equation}

When the extra dimension is large, instead of being rolled up to a small size, our spacetime can be  identified with some $4D$ hypersurface $y = $ const, which is orthogonal to the extra dimension. In this case the  metric induced in $4D$ is 
\begin{equation}
\label{Kramer-like solution}
ds^2 = f^a \; dt^2 - f^{- (a + b)}\; dr^2 - r^2 f^{(1 - a - b)}d\Omega^2.
\end{equation}
It is straightforward to verify that this line element, which in what follows we will call `Kramer-like',  is an exact solution to the field equation $^{(4)}R = 0$. 

In this section, we present a number of new families of analytic solutions of the form (\ref{the metric under study}) on the brane obtained by choosing 
\begin{equation}
\label{Choosing C}
C^2(r) = f^{1 - a - b},
\end{equation}
and fixing $A(r)$ or $B(r)$ as in the Kramer-like metric (\ref{Kramer-like solution}). Before discussing the new solutions, let us briefly examine some of the properties of this metric.

When $a = 0$, from (\ref{condition on a and b}) it follows that $b = \pm 1$ in which case (\ref{Kramer-like solution}) becomes 
\begin{equation}
\label{Spatial-Schw wormhole}
ds^2 = d t^2 - \left(1 - \frac{R_{0}}{R}\right)^{- 1} d R^2 - R^2 d\Omega^2,\;\;\;R_{0} \equiv = \frac{2 |b| m}{b}.
\end{equation}
For $R_{0} > 0$ this is the spatial-Schwarzschild wormhole \cite{Dadhich}, for $R_{0} < 0$ it is a naked singularity and for $R_{0} = 0$ it is Minkowski spacetime.
  
When $b = 0$, from (\ref{condition on a and b}) we find $a = \pm 1$ and the line element (\ref{Kramer-like solution}) reduces to 
\begin{equation}
\label{Schw solution with both signs}
ds^2 = \left(1 - \frac{2 \bar{m}}{R}\right)d t^2 - \left(1 - \frac{2 \bar{m}}{R}\right)^{- 1} d R^2 - R^2 d\Omega^2,\;\;\;\bar{m} \equiv  \frac{|a| m}{a}.
\end{equation}
For $\bar{m} > 0$ this is  
the Schwarzschild black hole  solution of general relativity with gravitational mass $M = \bar{m}$, a naked singularity for $\bar{m} < 0$ and Minkowski spacetime for $\bar{m} = 0$.

In any other case the gravitational mass is $M = a m$, which follows from the comparison of the asymptotic behavior $(r \rightarrow \infty)$ of $g_{tt}$ with Newton's theory [see (\ref{Post-Newtonian parameters}) and (\ref{Post-Newtonian parameters for Kramer's solution}) bellow].   To assure the positivity of $M$, in what follows  without loss of generality we take $m > 0$ and $a \geq 0$. Consequently, the  appropriate solution of  (\ref{condition on a and b}) is 
\begin{equation}
\label{a in terms of b}
a = - \frac{b}{2} + \frac{\sqrt{4 - 3 b^2}}{2} \geq 0,
\end{equation}
which holds in the range $- 2/\sqrt{3} \leq b \leq 1$. The Schwarzschild spacetime is recovered when $b = 0$. 

To study the singularities of the solutions it is useful to  calculate the Kretschmann scalar ${\cal{K}} = R_{\alpha\beta \gamma\delta}R^{\alpha\beta\gamma\delta}$. For the metric (\ref{the metric under study}) it is given by
\begin{equation}
\label{Kretschmann scalar}
{\cal{K}} = R_{\alpha \beta \mu\nu}R^{\alpha \beta \mu\nu} = 4 K_{1}^2 + 8 K_{2}^2 + 8 K_{3}^2 + 4 K_{4}^2, 
\end{equation}
where
\begin{eqnarray}
\label{K1, K2, K3, K4}
K_{1} &=& \frac{1}{ B^2}\left[\frac{A''}{A} - \frac{A' B' }{A B}\right], \nonumber \\
K_{2} &=& \frac{\left(C + r C'\right) A'}{r C B^2  A}, \nonumber \\
K_{3} &=& \frac{1}{B^2}\left(\frac{B' C'}{B C} + \frac{B'}{r B} - \frac{2 C'}{r C} - \frac{C''}{C}\right), \nonumber \\
K_{4} &=& \frac{\left(C + r C'\right)^2 - B^2}{r^2 C^2 B^2}. 
\end{eqnarray}
The finiteness of ${\cal{K}}$  is a necessary and sufficient criterion for the regularity of all curvature invariants. For the Kramer-like metric (\ref{Kramer-like solution}) we obtain
\begin{equation}
\label{Kretschmann scalar for Kramer-like metric}
{\cal{K}} = \frac{48 m^2 k}{r^8 f^{2(2 - a - b)}}, \;\;\;
\end{equation}
with
\begin{equation}
\label{k}
k = m^2 \left[2 (a + 1) + b - \frac{2 b^2 \left(a + 2\right)}{3} + \frac{a b^3}{6} + \frac{b^4 }{3}\right] - m r\left[2 \left(a + 1\right) +  b - \frac{b^2 \left(2 a + 3\right)}{3}\right] + \frac{r^2 \left(2 - b^2\right)}{2}.
\end{equation}

For $b = 0$, $(a = 1)$ the expression for $k$ reduces to $k = r^2 f^{2}$. Therefore, for $b = 0$ we recover the usual Schwarzschild singulatity at $r = 0$, viz., ${\cal{K}} = 48 m^2/r^6$, as expected. Since $(2 - a - b) > 0$, it follows that for $b \neq 0$ there is a physical singularity at $f = 0$, i.e., at $r = 2 m$.   

The physical radius of a sphere with coordinate $r$ is given by $R = r f^{(1 - a - b)/2}$. In the limit $r \rightarrow 2 m$ it  behaves either as $R \rightarrow 0$ or $R \rightarrow \infty$ depending on whether  $b \in \left[- 2/\sqrt{3}, 0\right)$ or $b \in \left(0, 1\right)$, respectively\footnote{The cases $b = 0$ and $b = 1$ corresponds to the Schwarzschild spacetime and to the spatial-Schwarszchild wormhole, respectively.}.  
 In the former range we have $\left(1 - a - b\right) > 0$, and consequently $R$ is a monotonically increasing function of $r$. In the latter range, where $\left(1 - a - b\right) < 0$, the physical radius $R$ is not a monotonic function of $r$; it reaches a minimum at $r = \bar{r} = m (1 + a + b) > 2 m$, viz., 
\begin{equation}
\label{Rmin}
\bar{R}= R(\bar{r}) =  m \sqrt{a b}\left(\frac{a + b + 1}{a + b - 1}\right)^{(a + b)/2},
\end{equation}
and then re-expands in $2 m < r < \bar{r}$ in such a way that $R \rightarrow \infty$ as $r \rightarrow 2 m$. We note that $2 m < \bar{r} < \left(1 + 2/\sqrt{3}\right) m \approx 2.155 m$ and $\bar{R} > 2 M$. Regarding $g_{tt}$ and $g_{rr}$,  we find that  $g_{tt} \rightarrow 0$ as $r \rightarrow 2 m$, in the whole range of $b$, i.e.,  $b \in \left[- 2\sqrt{3}, 1\right)$. In the same limit we have $g_{rr} \rightarrow \left(0, - 1, - \infty \right)$ for $-2/\sqrt{3}\leq b < - 1$, $b = -1$ and $- 1 < b < 1$, respectively. 

The effective energy density $\rho = T_{0}^{0}$ is given by 
\begin{equation}
\label{negative energy density}
8 \pi \rho = - \frac{m^2 a b}{r^4 f^{2 - a - b}}.
\end{equation}
Thus, for $b \neq 0$ the solutions have a naked singularity at $r = 2 m$ where $\rho$ diverges. 
For $b < 0$ the density is positive and  a traveler moving radially towards the center $R = 0$ reaches the singularity.  For $b > 0$ the density is negative and  a traveler moving  towards $r = 2m$ never reaches the singularity, instead (since there is a throat at $r = \bar{r}$, with physical radius $R = \bar{R}$) she/he  crosses into another space, which does not have a second flat  asymptotic because  $g_{tt} \rightarrow 0$ as $r \rightarrow 2 m$, despite the fact that $R \rightarrow \infty$ in this limit.

Consequently, even though   the spacetime configurations  with $b \in \left(0, 1\right)$  have a throat, and violate the null energy condition, they are topologically different  from wormholes which, by definition, connect two asymptotic regions.

\medskip

In order to experimentally distinguish between different asymptotically flat metrics, it is useful to calculate the post-Newtonian parameters $\beta$ and $\gamma$ in the Eddington-Robertson expansion \cite{Weinberg}
\begin{equation}
\label{Post-Newtonian parameters}
ds^2 = \left[1 - \frac{2 M}{R} + 2 (\beta - \gamma)\frac{M^2}{R^2} + \cdots\right]\; dt^2 - \left(1 + \frac{2 \gamma M}{R} + \cdots\right)\; dR^2 - R^2 d\Omega^2.
\end{equation}
The  parameter $\beta$ affects the precession of the perihelion and the Nordtvedt effect, while $\gamma $ affects the deflection of light and the time  delay of light \cite{Will}. 

For the solution (\ref{Kramer-like solution}) we obtain
\begin{eqnarray}
\label{Post-Newtonian parameters for Kramer's solution}
M &=& a m,\nonumber \\
\beta &=& 1,\nonumber \\
\gamma &=& 1 + \frac{b}{a}.
\end{eqnarray}
However, we should keep in mind that ``such hypothetical objects as braneworld black holes or wormholes, not necessarily of astrophysical size, need not necessarily conform to the restrictions on the post-Newtonian parameters obtained from the Solar system and binary pulsar observations, and it therefore makes sense to discuss the full range of parameters which are present in the solutions" \cite{Bronnikov 1}.

\subsection{Temporal Schwarzschild-Kramer-like solution} Following the same philosophy as in curvature coordinates, the line element  (\ref{Kramer-like solution}) can be used to generate other asymptotically flat vacuum solutions on the brane. For example, by demanding that 
\begin{equation}
\label{A and C for the temporal Schwarzschild-Kramer-like solution}
A^2(r) = f^a, \;\;\;C^2(r) = f^{1 - a - b},
\end{equation}
we find that the field equation (\ref{field equation for the general spherical metric}) has two solutions. One of them is just $B^2(r) = f^{- (a + b)}$, which gives back (\ref{Kramer-like solution}). The other one is a general solution which can be written as
\begin{equation}
\label{B for the temporal Schwarzschild-Kramer-like solution}
B^2(r) = \left[\frac{1 - \frac{m (2 + a + 2 b)}{2r}}{  1 - \frac{3\sigma m }{2 r}}\right] f^{- (a + b)}, 
\end{equation}
where $\sigma$ is a constant of integration. For an arbitrary $\sigma$ and $a = 1$ $(b = 0)$ we recover the temporal Schwarzschild metric (\ref{T-Schw exterior}). 

The total gravitational mass $M$ and the PPN parameters $\beta$ and $\gamma$ are given by 
\begin{eqnarray}
\label{Post-Newtonian parameters for temporal Kramer's solution}
M &=& a m,\nonumber \\
\beta &=& \frac{3 a - 2 b - 2 + 3 \sigma}{4 a},\nonumber \\
\gamma &=& \frac{3 a + 2 b - 2 + 3 \sigma}{4 a}.
\end{eqnarray}

If we denote 
\begin{equation}
\label{definition of r tilde and r0}
\tilde{r} = \frac{m (2 + a + 2 b)}{2},  \;\;\;\;r_{0} = \frac{3 \sigma m}{2},
\end{equation}
the solution can be written as 
\begin{equation}
\label{General temporal Schwarzschild-Kramer-like solution}
ds^2 = f^a d t^2 - \left(\frac{r - \tilde{r}}{r - r_{0}}\right)f^{- (a + b)} dr^2 - r^2 f^{1 - a - b} d\Omega^2.
\end{equation}

Here the Kretschmann scalar diverges at $r = 2 m$ and at $r = \tilde{r}$. In contrast, $r = r_{0}$ is a coordinate singularity; not a physical one. It should be noted that $0 < \tilde{r} < 2 m$ 
for any $a > 0$. Therefore, the condition $g_{rr} < 0$ implies that the above solution makes sense only for $r \geq r_{0} = 3 m \sigma /2$. 
If $r_{0} \leq 2m$, i.e. $\sigma \leq 4/3$,  the solution is a naked singularity.
However for $r_{0} > 2 m$ $(\sigma > 4/3)$ it is a  traversable wormhole. 

Since the Kretschmann scalar is regular at $r _{0}$, the singularity can be removed by introducing a new coordinate $x$ by the relation $r = r_{0} + x^2$. The explicit form of the solution in terms of $x$ is 

\begin{equation}
\label{General temporal Schwarzschild-Kramer-like solution in terms of x}
ds^2 = \left(\frac{x^2 + r_{0} - 2 m}{r_{0} + x^2}\right)^a d t^2 - \frac{4 \left(r_{0} + x^2\right)^{a + b}\left(x^2 + r_{0}  -
 \tilde{r}\right)}{\left(x^2 + r_{0} - 2m\right)^{a + b}} \; dx^2 - \frac{\left(x^2 + r_{0}\right)^{a + b + 1}}{\left(x^2 + r_{0} - 2m\right)^{a + b  - 1}} d\Omega^2.
\end{equation}
For $r_{0} > 2 m$ $(\sigma > 4/3)$, the metric is regular for all values of $x \in \left( - \infty , + \infty \right)$ and invariant under sign reversal $x \rightarrow - x$. Therefore, both $x \rightarrow \infty$ and $x \rightarrow - \infty$ are flat asymptotics. The physical radius $R$ of a spherical shell with coordinate $x$ is
\begin{equation}
\label{physical radius of a spherical shell with coordinate x}
R(x) =  \frac{\left({{x}}^2 + r_{0}\right)^{(a + b + 1)/2}}{\left({{x}}^2 + r_{0} - 2 m\right)^{(a + b - 1)/2}}. 
\end{equation}

 The equation $dR/d{x} = 0$ has the following roots
\[
{{x}}_{0} = 0, \;\;\;{{x}}_{(\pm)} = \pm \sqrt{\bar{r} - r_{0}}, \;\;\;\; 
\]
where 
\[
\bar{r} = m \left(1 + a + b\right).
\]
 The ${{x}}_{(\pm)}$ solutions are real only if $\bar{r} > r_{0}$. Since $r_{0} > 2 m$,  this imposes the condition $(a + b ) > 1$, which  requires $b \in \left(0, 1\right)$. Thus, the metric (\ref{General temporal Schwarzschild-Kramer-like solution in terms of x}) is regular for all values of $x$ and $R(x)$ has (real) extremuma at $x = x_{(\pm)}$  if
\begin{equation}
\label{condition for minimum at x neq 0}
2 m < r_{0} < \bar{r} < \left(1 + \frac{2}{\sqrt{3}}\right) m \approx 2.155 m. 
\end{equation}
In terms of the dimensionless quantities $\sigma = 2 r_{0}/3 m$ and  $\bar{\sigma} = 2 \bar{r}/3 m$, this inequality can be written as\footnote{We note that $\bar{\sigma}$ is bounded from above, namely,    $\bar{\sigma} < {\bar{\sigma}}_{max} = 2\left(3 + 2\sqrt{3}\right)/9 \approx 1.436$, which  corresponds to $b = 1/\sqrt{3} \approx 0.577$. } 
\begin{equation}
\label{condition for minimum at x neq 0 in terms of sigma}
\frac{4}{3} < \sigma < \bar{\sigma} < \frac{2\left(3 + 2\sqrt{3}\right)}{9} \approx 1.436.
\end{equation}
In Table $1$ we provide  $\bar{\sigma} = 2 (1 + a + b)/3$ calculated for various values of $b \in \left(0, 1\right)$. 
\begin{center}
\begin{tabular}{|c|c|c|c|c|c|c|c|} \hline
\multicolumn{8}{|c|}{\bf Table 1. $\bar{\sigma}$ for  various values of $b \in \left(0, 1\right)$} \\ \hline
 \multicolumn {1}{|c|}{$b$} & 
$0.01$ &  $0.1$ &$0.3$&$0.5$&$0.8$&$0.9$&\multicolumn{1}{|c|}{$0.99$} \\ \hline\hline
$\bar{\sigma}$ & $1.337$& $1.364$& $1.410$&$1.434$ &$1.414$&$1.384$&$1.340$ \\ \hline
\end{tabular}
\end{center}
In summary, we have the following cases:

\paragraph{Case 1:}  The function $R = R({x})$ has only one extremum, which is located  at $\bar{x} = 0$ if (i) $b \in \left[- 2/\sqrt{3}, 0\right) $  and {\it any} $r_{0} > 2 m$ $(\sigma > 4/3)$, or (ii) $b \in \left(0, 1\right)$ and  $r_{0} > 3 m \bar{\sigma}/2$ (for $b > 0$, a  sufficient criterion for one extremum is  $r_{0} > 2.155 m$). Under these conditions $\bar{x} = 0$ is the minimum of (\ref{physical radius of a spherical shell with coordinate x}). Therefore, in this case  the  wormhole throat is located at $R = R_{0}$ given by

\begin{equation}
\label{location of the throat for temporal Schw. wormhole}
R_{0} = \frac{3 \sigma M}{2 a}\left(1 - \frac{4}{3 \sigma}\right)^{(1 - a - b)/2}, 
\end{equation}
with $(b < 0, \sigma > 4/3)$ and $(b > 0, \sigma > \bar{\sigma})$. We observe that the condition $\sigma > \bar{\sigma}$ is automatically satisfied when $b < 0$, because ${\bar{\sigma}}_{(b < 0)} < 4/3$,   while on the contrary  ${\bar{\sigma}}_{(b > 0)} > 4/3$. 

The metric obtained by Casadio, Fabbri and Mazzacurati \cite{CFM} for new braneworld black holes and the symmetric traversable wormhole solutions discussed by Bronnikov and Kim \cite{Bronnikov 1}, in their example $2$, are  restored from (\ref{General temporal Schwarzschild-Kramer-like solution in terms of x})-(\ref{location of the throat for temporal Schw. wormhole}) in the case $a = 1$, $b = 0$, which in turn reduces to the Schwarzschild  metric for $\sigma = 1$, i.e. $r_{0} = \tilde{r} = 3 m/2$. 

From (\ref{location of the throat for temporal Schw. wormhole}) we find that in this case we can have  $R_{0} < 2 M$ or $R_{0} > 2 M$ depending on whether $b < 0$ or $b \geq 0$.  To illustrate this  we have evaluated (\ref{location of the throat for temporal Schw. wormhole}) for some specific values of $b$ and $\sigma$. The results are presented in Figure $1$.

For a light signal the time for propagation, measured by an external observer,  from the throat $x = 0$ to some $x = x_{0} > 0$ is given by 

\begin{equation}
\label{time to the throat}
\Delta t =  2 m \int^{{\bar{x}}_{0}}_0\sqrt{{\bar{x}}^2 + \frac{3 \sigma}{2} - \frac{2 + a + 2 b}{2}}\left[1 - \frac{2}{{\bar{x}}^2 + 3 \sigma/2}\right]^{- (a + b/2)} d\bar{x},
\end{equation}
where $\bar{x}$ is a dimensionless quantity defined as $\bar{x} = x/\sqrt{m}$. To test the above equations, let us take, e.g.,  $b = - 0.2$ and $\sigma = 1.5$, which are in the range for which (\ref{location of the throat for temporal Schw. wormhole}) defines the location of the throat. For these values $R_{0} \approx 1.83 M$ and $\Delta t$ is finite for any ${\bar{x}}_{0} < \infty$. One can numerically verify that $\Delta t$ is finite for any $b < 0$ and $\sigma > 4/3$. The conclusion is that for a distant observer a light signal crosses the Schwarzschild radius $R = 2 M$, which now is not a horizon, in a finite time.

 Also, from (\ref{location of the throat for temporal Schw. wormhole})  we find
\begin{equation}
\label{dR/dsigma}
\frac{1}{R_{0}}\frac{d R_{0}}{d \sigma} = \frac{\sigma - \bar{\sigma}}{\sigma \left(3 \sigma - 4\right)}.
\end{equation}
Thus, in the present case $\left(\sigma > \bar{\sigma}\right)$,  $R_{0}$ increases monotonically with $\sigma.$ 

We note that at $x = 0$  

\begin{eqnarray}
\label{energy conditions}
\rho > 0, \;\;\;&&\rho - p_{r} > 0, \;\;\;\rho + p_{r} \sim \frac{\bar{\sigma} - \sigma}{r_{0} - \tilde{r}},\nonumber \\
 &&\rho + p_{\perp} > 0, \;\;\;\rho - p_{\perp} \sim \frac{\bar{\sigma} - \sigma}{r_{0} - \tilde{r}}.
\end{eqnarray}
where $T_{0}^{0} = \rho$, $T_{1}^{1} = - p_{r}$, $T_{2}^{2} = - p_{\perp}$. Thus, in the present case $\left(\sigma > \bar{\sigma}\right)$, in a neighborhood of the wormhole throat the effective EMT violates not only the null energy condition $(\rho + p_{r}) > 0$, which is in agreement with a well known general result, but also violates the dominant energy condition.

\begin{figure}[tbp] 
  \centering
  \includegraphics[bb=0 0 991 991,width=3.00in,height=3.00in,keepaspectratio]{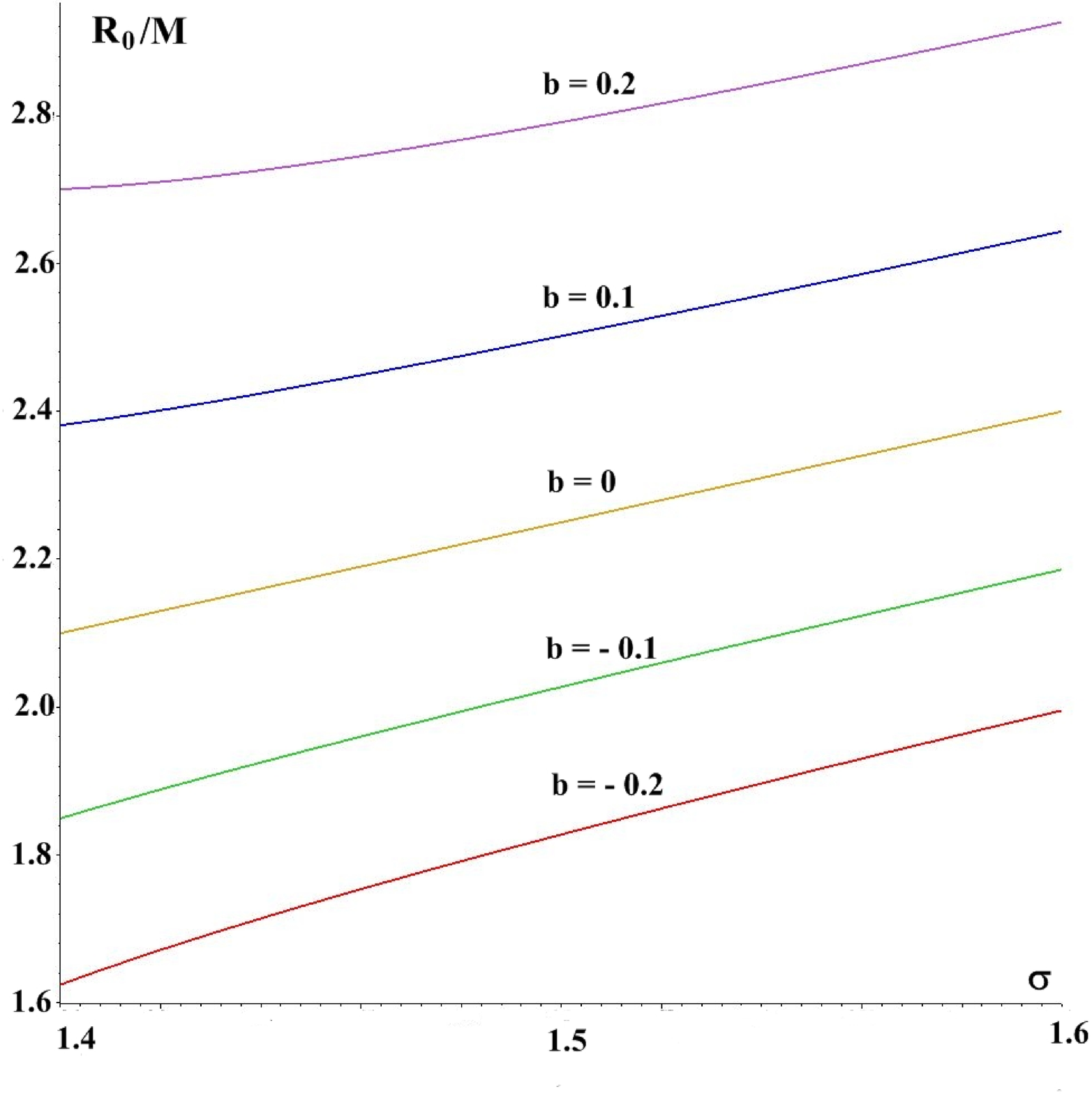}
  \caption{In Case $1$ the wormhole throat is located at $R = R_{0}$. The figure shows that $R_{0}$ increases with $\sigma$. More interesting is that for $b < 0$ there is a range of values of $\sigma $ for which $R_{0} < 2 M$. For $b > 0$, $R_{0} > 2 M$.}
  \label{Figure $1$}
\end{figure}

\begin{figure}[tbp] 
  \centering
  \includegraphics[bb=0 0 990 990,width=3.00in,height=3.00in,keepaspectratio]{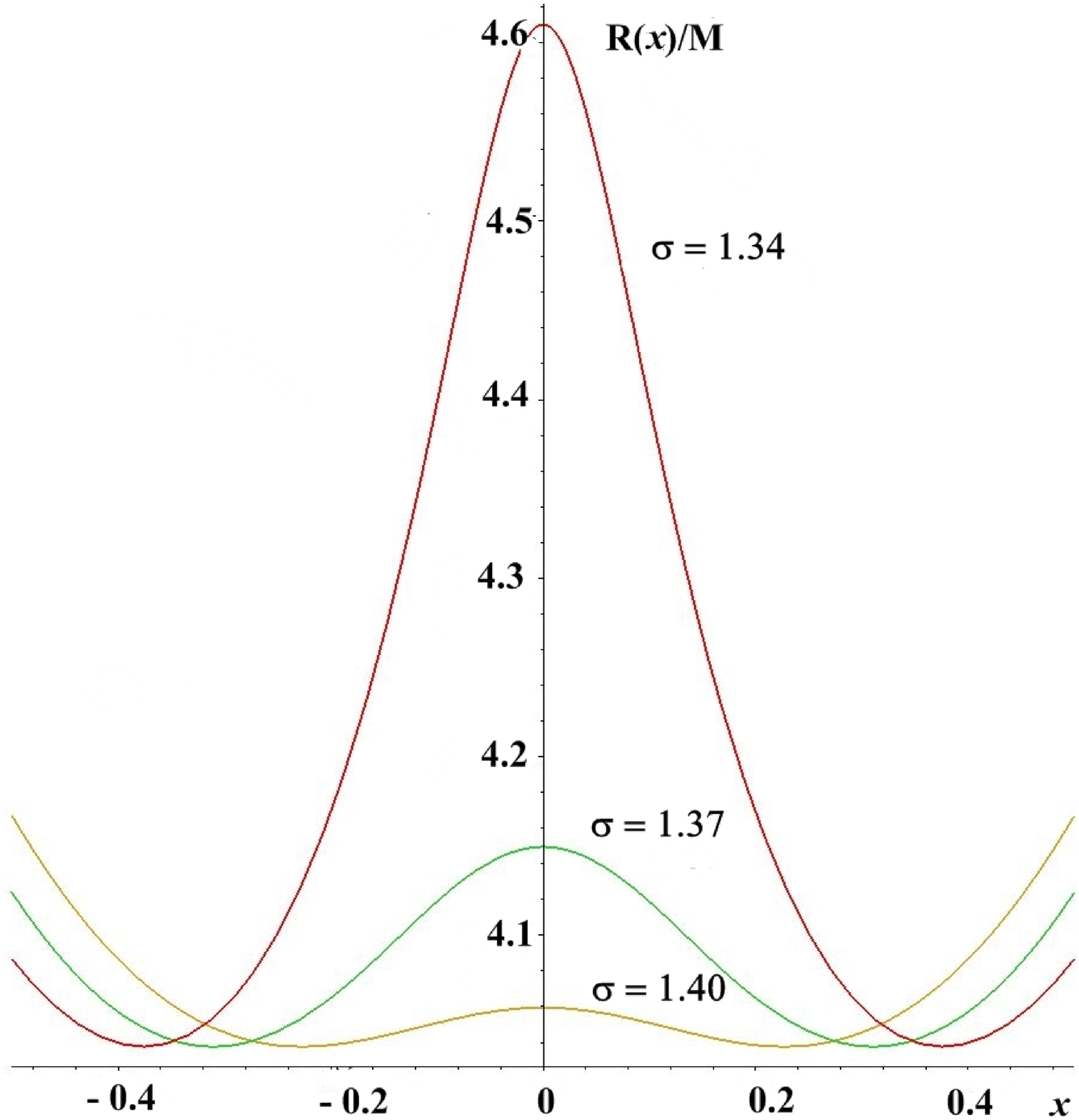}
  \caption{In Case $2$ there are three extremuma, i.e., the spacelike surfaces $R_{(x_{\pm})}$ and $R_{0}$ are turning points for light and all material particles.  The figure gives the physical radius as a function of $x$ for $b = 0.5$ and the values of $\sigma$ considered in Table $2$. Without loss of generality we have set $m = 1$, which is equivalent to introducing a dimensionless coordinate $x \rightarrow x/\sqrt{m}$ in $(30)$. As we approach $r = 2 m$ we never reach distances $R < R_{(x_{\pm})} = 4.038 M$ regardless of the choice of $\sigma$, although $R_{(x_{\pm})}$ is attained at different values of $x$.  After a re-bounce at $R_{0}$, which does depend on $\sigma$ but corresponds to the same $x$, namely $x = 0$,  we cross into another space which is the double of the one we started in. The two minimuma coalesce for $\sigma \rightarrow \bar{\sigma}$}
  \label{Figure $2$}
\end{figure}

\paragraph{Case 2:} When $b \in \left(0, 1\right)$ and $2 m < r_{0} < \bar{r}$, or what is equivalent $4/3 < \sigma < \bar{\sigma}$ ,  there are three extremuma. One of them is at $x = 0$ and corresponds to $R = R_{0}$ given by (\ref{location of the throat for temporal Schw. wormhole}); the other two are at $x = x_{\pm}$ for which 

\begin{equation}
\label{Rmin for b positive, temporal Schwarzschild-Kramer-like}
R_{(x_{\pm})} = M \sqrt{\frac{b}{a}}   \left(\frac{a + b + 1}{a + b - 1}\right)^{(a + b)/2} > 2 M. 
\end{equation}

In this case $\left(1 - a - b\right) < 0$, therefore from (\ref{location of the throat for temporal Schw. wormhole}) it follows that $R_{0} \rightarrow \infty$ as $\sigma \rightarrow \left(4/3\right)^{+}$. In addition, $R_{0} \rightarrow  R_{(x_{\pm})}^{+}$ as $\sigma \rightarrow \bar{\sigma}$, i.e. $R_{0}$ decreases with the increase of $\sigma$,  which is in agreement with (\ref{dR/dsigma}) because in the present case $\sigma < \bar{\sigma}$. What this means is that   
in this case $R_{(x_{\pm})}$ is a local minimum and $R_{0}$ is a local maximum. This is illustrated in Figure $2$. 

As an example,  let us choose  some particular $b$ in the range  $\left(0, 1\right)$, e.g. $b = 0.5$. For this choice $R_{(x_{\pm})} = 4.038 M$. Besides, from Table $1$ we get  $\bar{\sigma} = 1.434$, $(\bar{r} = 2.151 m)$. In Table $2$ we compute $R_{0}$ for  various values of $\sigma$ in the range $1.333 < \sigma < 1.434$ $(2 m < r_{0} < 2.151 m)$. Similar results, illustrating the fact that $R_{0} > R_{(x_{\pm})} > 2 M$,  can be obtained for other values of $b$ and $\bar{\sigma}$ considered in Table $1$.  
\begin{center}
\begin{tabular}{|c|c|c|c|c|} \hline
\multicolumn{5}{|c|}{\bf Table 2. $R_{0}/M$ for $b = 0.5$ and $\sigma \in \left(4/3, 1.434\right)$} \\ \hline
 \multicolumn {1}{|c|}{$\sigma$} & 
 $1.34$&$1.37$&$1.40$&\multicolumn{1}{|c|}{$1.43$} \\ \hline\hline
$R_{0}/M$ & $4.610$ &$4.149$&$4.059$&$4.037$ \\ \hline
\end{tabular}
\end{center}

From (\ref{energy conditions}) it follows that, in the present case $(\bar{\sigma} > \sigma)$, the effective EMT satisfies the weak, null and dominant energy conditions in a neighborhood of $x = 0$. However, using (\ref{condition for minimum at x neq 0 in terms of sigma}) and Table $1$, it is not difficult to verify that these conditions are now violated  
in a neighborhood of $x = x_{(\pm)}$, which is where in this case the wormhole throat is located at, as expected. 
\subsection{Spatial Schwarzschild-Kramer-like solution} In order to generalize the spatial-Schwarzschild metric (\ref{S-Schw exterior}) we now choose
\begin{equation}
\label{B and C for Spatial Kramer-Schwarzschild-like solution}
B^2(r) = \left(1 - \frac{2 \alpha m}{r}\right)^{- (a + b)}, \;\;\;C^2(r) = \left(1 - \frac{2 \alpha m}{r}\right)^{1 - a - b},
\end{equation}
where $\alpha$ is some positive constant. Substituting these expressions into (\ref{field equation for the general spherical metric}) we obtain a second order differential equation for $A(r)$. To determine the arbitrary constants of integration we impose two conditions. First,  that $A \rightarrow 1$ as $r \rightarrow \infty$. Second, that for $a = 1$ $(b = 0)$ we recover the spatial Schwarzschild solution. With these conditions we obtain
\begin{equation}
\label{A for Spatial Kramer-Schwarzschild-like solution}
A^2(r) = \frac{1}{\alpha^2}\left[\alpha - 1 + \left({1 - \frac{2 \alpha m}{r}}\right)^{(a - b)/2}\right]^{2}\left(1 - \frac{2 \alpha m}{r}\right)^b.
\end{equation}
For this solution the total gravitational mass and the PPN parameters are
\begin{eqnarray}
\label{Post-Newtonian parameters for the spatial Schwarzschild-Kramer-like solution}
M &=& m \left[a + b(\alpha - 1)\right],\nonumber \\
\beta &=& \frac{1}{2}\left\{1 + \frac{\alpha \left[a^2 + \alpha b^2 (\alpha - 1)\right]}{\left[a + b \left(\alpha - 1\right)\right]^2}\right\},\nonumber \\
\gamma &=& \frac{\alpha (a + b)}{a + b (\alpha - 1)}.
\end{eqnarray}

The above solution for $\alpha = 1$ reduces to the Kramer-like solution (\ref{Kramer-like solution}), and for $\alpha \neq 1$, but $a = 1$ $(b = 0)$,   gives back  (\ref{S-Schw exterior}), as expected.  
In general, for $\alpha \neq 1$, $a \neq 1$ and $b \neq 0$ the quantities $K_{1}$ and $K_{2}$ in (\ref{K1, K2, K3, K4}) behave like $\sim \left[1/(r^2 Af^{2 - a - b})\right]$. Therefore the Kretschmann scalar diverges at $r = 0$, $A = 0$ and $r = 2 m \alpha$, as $\left(2 - a - b\right) > 0$ in the whole range of $b$. 

However, there are two cases in which the curvature invariants are regular at $r = 2 m \alpha$. One of them is $a = 1$, $b = 0$ discussed by Casadio, Fabbri and Mazzacurati \cite{CFM}, the other case is when $a = 0$ and $b = 1$. In the latter case the metric (\ref{B and C for Spatial Kramer-Schwarzschild-like solution})-(\ref{A for Spatial Kramer-Schwarzschild-like solution}) becomes

\begin{equation}
\label{a = 0, b = 1, Spatial Kramer-Schwarzschild-like solution}
ds^2 = \frac{1}{\alpha^2}\left[1 + \left(\alpha - 1\right)\sqrt{1 - \frac{2 \alpha m }{r}}\right]^2 d t^2 - \left(1 - \frac{2 \alpha m}{r}\right)^{- 1} d r^2 - r^2 d\Omega^2.
\end{equation}
For $\alpha = 1$ we recover the spatial-Schwarzschild wormhole (\ref{Spatial-Schw wormhole}) with $R_{0} = 2 \alpha m$. For $\alpha < 1$, the equation $g_{tt} = 0$ has no positive solutions for $r$. Thus, there is no horizon. Since the Kretschmann scalar is regular at $r = r_{0} = 2 \alpha m$,  the metric is regularized at $r = r_{0}$ by the substitution $x^2 = r - r_{0}$. As a result, (\ref{a = 0, b = 1, Spatial Kramer-Schwarzschild-like solution}) becomes

\begin{equation}
\label{a = 0, b = 1, Spatial Kramer-Schwarzschild-like solution in terms of x}
ds^2 = \frac{1}{\alpha^2}\left[1 + \frac{(\alpha - 1) |x|}{\sqrt{x^2 + r_{0}}}\right]^2 d t^2 - 4 \left(x^2 + r_{0}\right)d x^2 - (x^2 + r_{0})^2 d\Omega^2,\;\;\;\alpha > 1,
\end{equation}
which is a symmetric traversable wormhole with throat at  $R_{0} = 2 M \alpha/(\alpha - 1)$ and total gravitational mass $M = m \left(\alpha - 1\right)$. We emphasize that here $R_{0} > 2 M$ for all values of $\alpha$.

 In order to make contact with other works in the literature, let us notice that the solution for $\alpha = 1$ can alternatively be written as 
\begin{eqnarray}
\label{solution for alpha = 1}
d s^2 &=& \left[\kappa + \lambda f^{(a - b)/2}\right]^2 f^b d t^2 - f^{- (a + b)} d r^2 - r^2 f^{1 - a - b} d\Omega, \nonumber \\
\end{eqnarray}
where $\kappa$ and $\lambda$ are arbitrary constants. The choice  $\kappa = 0$ and $\lambda = 1$ gives back the Kramer-like solution (\ref{Kramer-like solution}); $\lambda = 0$ and $\kappa = 1$ yields a line element which is similar to (\ref{Kramer-like solution}) with $a$ replaced by $b$ and vice versa. The class of self-dual Lorentzian wormholes discussed in Dadhich {\it et al} \cite{Dadhich}, which in our notation are given by (\ref{a = 0, b = 1, Spatial Kramer-Schwarzschild-like solution in terms of x}),  is  recovered from (\ref{solution for alpha = 1}) in the special cases where  $a = 1$, $b = 0$ or  $a = 0$, $b = 1$. This class of solutions is a particular case of the wormhole spacetimes discussed by Bronnikov and Kim \cite{Bronnikov 1} in their example $4$. 

\medskip 

$\bullet$ A more detailed investigation shows that in the case where $\alpha = 1$, there is a family of ``spatial Kramer-like solutions" not included in (\ref{solution for alpha = 1}). Indeed, it is not difficult to verify that the line element 

\begin{equation}
\label{spatial Kramer-like solution}
ds^2 = f^{[a + b + \sqrt{1 - 3 a b}]/2} \; dt^2 - f^{- (a + b)}\; dr^2 - r^2 f^{(1 - a - b)}\; d\Omega^2,
\end{equation}
also satisfies the field equation (\ref{field equation for the general spherical metric}). 
Note that $a + b + \sqrt{1 - 3 a b} > 0$ for all $a > 0$ (i.e., $- 2\sqrt{3} \leq  b < 1$).  The Schwarzwschild geometry can be recovered in two distinct limits:  either $a = 1$, $b = 0$ or $a = 0$, $b = 1$.  From an observational point of view this solution is distinct from the ones considered above. This follows from the fact that the PPN parameters are different from the ones calculated in (\ref{Post-Newtonian parameters for the spatial Schwarzschild-Kramer-like solution}). Namely, for the line element (\ref{spatial Kramer-like solution}) we find
\begin{eqnarray}
\label{Post-Newtonian parameters for spatial Kramer solution}
M &=& \frac{m}{2}\left(a + b + \sqrt{1 - 3 a b}\right),\nonumber \\
\beta &=& 1,\nonumber \\
\gamma &=& \frac{2(a + b)}{a + b + \sqrt{1 - 3 a b}}.
\end{eqnarray}

$\bullet$ It should be  emphasized that we can use any of the above solutions to generate other asymptotically flat 
solutions to (\ref{field equation for the general spherical metric}) that contain the Schwarzschild spacetime. As an example, let us consider the case where  the temporal part of the metric is identical to $g_{tt}$ in the spatial Kramer-like solution (\ref{spatial Kramer-like solution}).  From (\ref{field equation for the general spherical metric}) we obtain a first-order differential equation for $B(r)$, which can be easily integrated. The new solution can be written as 
\begin{equation}
\label{spatial Kramer-like solution modified}
ds^2 = f^{[a + b + \sqrt{1 - 3 a b}]/2} \; dt^2 - \left(\frac{r - \hat{r}}{r - r_{0}}\right) f^{- (a + b)}\; dr^2 - r^2 f^{(1 - a - b)}\; d\Omega^2,
\end{equation}
with
\[
\hat{r} = \frac{m \left[4 + 3 \left(a + b\right) - \sqrt{1 - 3 a b}\right]}{4}, \;\;\;\;r_{0} = \frac{3 m \sigma}{2} = \mbox{constant},
\]
where $\sigma$ is a dimensionless, arbitrary,  constant of integration. For $r_{0} = \hat{r}$ the metric (\ref{spatial Kramer-like solution modified}) reduces to (\ref{spatial Kramer-like solution}). In addition, for $a = 1$, $b = 0$ the solutions (\ref{General temporal Schwarzschild-Kramer-like solution}) and (\ref{spatial Kramer-like solution modified}) yield the temporal Schwarzschild  metric (\ref{T-Schw exterior}).

It should be noted that $0 < \hat{r} < 2 m$ in the whole range $- 2/\sqrt{3} \leq b < 1$. The Kretschmann scalar diverges at $r = 0$ and $r = 2 m$,  but is regular at $r = \hat{r}$ and $r = r_{0}$. Once again, smooth continuation at $r = r_{0} > 2 m$, which requires $\sigma > 4/3$, is achieved in terms of the coordinate $x$ defined by the relation $r = r_{0} + x^2$, viz.,
\begin{equation}
\label{spatial Kramer-like solution modified in terms of x}
ds^2 = \left(\frac{x^2 + r_{0} - 2 m}{x^2 + r_{0}}\right)^{[a + b + \sqrt{1 - 3 a b}]/2} \; dt^2 - 
\frac{4 \left(x^2 + r_{0} - \hat{r}\right)\left(x^2 + r_{0}\right)^{a + b}}{\left(x^2 + r_{0} - 2 m\right)^{a + b}}\; dx^2 - 
\frac{\left(x^2 + r_{0}\right)^{a + b + 1}}{\left(x^2 + r_{0} - 2 m\right)^{a + b - 1}}\; d\Omega^2.
\end{equation}
Here the physical radius of a spherical shell with coordinate $x$ is the same as in (\ref{physical radius of a spherical shell with coordinate x}). Therefore, under the same conditions on $b$ and $\sigma$ we have the Cases $1$ and $2$ discussed above. The only difference is that now
the total gravitational mass is given by the first equation in (\ref{Post-Newtonian parameters for spatial Kramer solution}). Therefore, in Case $1$ the wormhole throat is at 
\begin{equation}
\label{location of the throat for the spatial Kramer-like solution modified}
R_{0} = \frac{3 \sigma M}{a + b + \sqrt{1 - 3 a b}}\left(1 - \frac{4}{3 \sigma}\right)^{(1 - a - b)/2},  \;\;\;\sigma > 4/3.
\end{equation}
In case $2$ we now have 
\begin{equation}
\label{R + - for the spatial Kramer-like solution modified}
R_{(x_{\pm})} = \frac{2 M \sqrt{a b}}{a + b + \sqrt{1 - 3 a b}}\left(\frac{a + b + 1}{a + b - 1}\right)^{(a + b)/2}.
\end{equation}
One can verify that in Case $2$, the weak, null and dominant energy conditions are satisfied in a neighborhood of $R_{0}$.  
Once again as in (\ref{location of the throat for temporal Schw. wormhole}) and (\ref{Rmin for b positive, temporal Schwarzschild-Kramer-like}), (i) if $b < 0$ we can have $R_{0} < 2 M$ for a wide range of values of $\sigma$ (see Fig. $1$); (ii) $R_{0} > 2 M$ for $b > 0$ (with $\sigma > \bar{\sigma}$), and (iii) $R_{(x_{\pm})} > 2 M$ for all $b > 0$.

\section{Symmetric wormholes from solutions with one regular asymptotic}

Let us note that, from a mathematical point of view,  the Kramer-like metric (\ref{Kramer-like solution}) and its temporal Schwarzschild generalization (\ref{General temporal Schwarzschild-Kramer-like solution}) differ only by a factor   in $g_{rr}$.  However, from a physical point of view they are very different: these metrics  have distinct PPN parameters and  (\ref{Kramer-like solution}) only has one asymptotic region  for $b > 0$. On the other hand,  (\ref{General temporal Schwarzschild-Kramer-like solution}) describes symmetric wormholes in the whole range of $b$ provided $r_{0} > 2 m$. A similar situation occurs between the spatial Kramer-like solution (\ref{spatial Kramer-like solution}) and its generalization     (\ref{spatial Kramer-like solution modified}). The aim of this section is to  generate two more families of traversable symmetric wormholes from solutions that  only have one regular asymptotic region. 

The new solutions arise from the observation that any solution to the field equation (\ref{the metric under study}) originates new ones given by   
\begin{equation}
\label{old solutions originate new solutions}
ds^2 = A^2(r) dt^2 - h(r) B^2(r) dr^2 - r^2 C^2(r) \; d\Omega^2, 
\end{equation}
 where 
\begin{equation}
\label{equation for h}
h(r) = \left[1 + c\; e^{- \int{\frac{B^2\; dr}{L^2\left(L'/L + A'/2 A\right)}}}\right]^{- 1}; 
\end{equation}
$L \equiv r C(r)$,  and $c$ is some arbitrary constant. 

\medskip

$\bullet$ First, let us consider the solution (\ref{General temporal Schwarzschild-Kramer-like solution}). If we demand that it must contain the Schwarzschild spacetime for $b = 0$ $(a = 1)$, then  we should set  
$\sigma = 1$. The result is the ``temporal Kramer-like" metric 
\begin{equation}
\label{temporal Kramer-like solution}
ds^2 = f^a \; dt^2 - \frac{2 (f + 1) + (f - 1)(a + 2 b)}{1 + 3 f}f^{- (a + b)}\; dr^2 - r^2 f^{(1 - a - b)}d \Omega^2.
\end{equation}
We note that $g_{rr} < 0$ for any positive $a$. If $a = 0$ and  $b = 1$ $(b = - 1)$, the solution describes a wormhole with throat at $R_{0} = 3 m/2$ $(R_{0} = - m/2, \; m < 0)$. In general, for any  $b < 0$ there is naked singularity at $r = 2m$. However, for $b > 0$ the physical radius has  a minimum  at $\bar{r} = m \left(1 + a + b\right) > 2 m$. 

Substituting  (\ref{temporal Kramer-like solution}) into (\ref{equation for h}) we generate another solution to (\ref{field equation for the general spherical metric}), namely
 \begin{equation}
\label{temporal Kramer-like solution, wormhole}
ds^2 = f^a \; dt^2 - \left(\frac{r - 3 m/2}{r - r_{0}}\right)\left[\frac{2 (f + 1) + (f - 1)(a + 2 b)}{1 + 3 f}\right]f^{- (a + b)}\; dr^2 - r^2 f^{(1 - a - b)}d \Omega^2,
\end{equation}
where $r_{0} \equiv \left(3 m + c\right)/2$.

\medskip

$\bullet$ Second, we  consider the line element (\ref{spatial Kramer-like solution modified}). If we set  $\sigma = 1$ we recover the Schwarzschild metric when  $b = 0$ $(a = 1)$. With this choice the solution becomes
\begin{equation}
\label{mixed Kramer-like solution}
ds^2 = f^{[a + b + \sqrt{1 - 3 a b}]/2} \; dt^2 - \frac{4 \left(f + 1\right) + \left(f - 1\right)\left[3\left(a + b\right) - \sqrt{1 - 3 a b}\right]}{2 \left(1 + 3 f\right)}               \;f^{- (a + b)}\; dr^2 - r^2 f^{(1 - a - b)} d\Omega^2.
\end{equation}
For this metric we find
\begin{eqnarray}
\label{Post-Newtonian parameters for mixed Kramer solution}
M &=& \frac{m}{2}\left(a + b + \sqrt{1 - 3 a b}\right),\nonumber \\
\beta &=&  \frac{2 + a + b + 5 \sqrt{1 - 3 a b}}{4(a + b + \sqrt{1 - 3 a b})},\nonumber \\
\gamma &=& \frac{5(a + b) + 2 + \sqrt{1 - 3 a b}}{4\left(a + b + \sqrt{1 - 3 a b}\right)}.
\end{eqnarray}
Substituting (\ref{mixed Kramer-like solution}) into (\ref{equation for h}) we get another  solution to (\ref{field equation for the general spherical metric}), viz., 
\begin{equation}
\label{mixed Kramer-like solution, wormhole}
ds^2 = f^{[a + b + \sqrt{1 - 3 a b}]/2} \; dt^2 - \left(\frac{r - 3 m/2}{r - r_{0}}\right)\left[\frac{4 \left(f + 1\right) + \left(f - 1\right)\left[3\left(a + b\right) - \sqrt{1 - 3 a b}\right]}{2 \left(1 + 3 f\right)}\right]               \;f^{- (a + b)}\; dr^2 - r^2 f^{(1 - a - b)} d\Omega^2.
\end{equation}
Thus, although the original metrics (\ref{temporal Kramer-like solution}) and (\ref{mixed Kramer-like solution}) only have one regular asymptotic region for $b >  0$,  the new solutions (\ref{temporal Kramer-like solution, wormhole}) and (\ref{mixed Kramer-like solution, wormhole}) in terms of the coordinate $x$ defined by  $r = x^2 + r_{0}$ are regular and symmetric for all values of $b$ and $r_{0} > 2 m$, so that both $x \rightarrow \infty$ and $x \rightarrow - \infty$ are flat asymptotics. All our symmetric wormholes have a factor proportional to $\left(r - r_{0}\right)^{- 1}$ in $g_{rr}$. In this regard, it is interesting to mention  a general result obtained by Bronnikov and Kim \cite{Bronnikov 1} in curvature coordinates $(g_{\theta\theta} = -R^2)$. They showed that in traversable, twice asymptotically flat, wormhole solutions the metric function $g_{RR}$ near the throat must behave as $\left(R - R_{0}\right)^{- 1}$.

\section{Summary}

The aim of this work has been to generate new exact static, spherically symmetric Lorentzian wormhole solutions on the brane. Since (\ref{field equation for the general spherical metric}) is a second-order differential equation for $A(r)$ and $C(r)$ and first order for $B(r)$, the simplest way for generating static solutions is to provide some smooth functions $A(r)$ and $C(r)$. Then, the field equation  (\ref{field equation for the general spherical metric}) reduces to a linear first-order differential equation for $B(r)$.  In the context of curvature coordinates, $C(r) = 1$, Bronnikov and Kim \cite{Bronnikov 1} have given a  thorough discussion of the general  conditions on the so-called redshift function $\ln{A(r)}$ under which the solution describes symmetric and asymmetric wormholes.

In this paper, to accomplish our goal we have solved (\ref{field equation for the general spherical metric}) using a different approach: (i) we have relaxed  the condition $C(r) = 1$, which is  required  in curvature coordinates. Instead we have chosen $C(r) = f^{(1 - a - b)/2}$. From a physical point of view, this choice automatically incorporates the requirement of existence of a throat. Namely, $R(r)= r f^{(1 - a - b)/2}$ has at least one regular minimum (a throat) at some $r = \bar{r} > 2 m$,  for $b > 0$ (see (\ref{Rmin})). 
In principle, one can always set $C = 1$ by redefining the radial coordinate. However,  the line element (\ref{Kramer-like solution})  cannot in general (i.e., for any $b \neq 0$) be written in a simple analytical form in terms of the radial coordinate $R = r f^{(1 - a - b)/2}$. Therefore, from a practical point of view the choice  $C(r) = f^{(1 - a - b)/2}$ generates solutions to (\ref{field equation for the general spherical metric}) which are algebraically simple in terms of $r$, but (in general) not expressible in terms of elementary functions of $R$; (ii) we have demanded that the spacetime must contain the Kramer-like spacetime (\ref{Kramer-like solution}), which is a vacuum solution on the brane constructed from the Kaluza-Klein $5D$ solution (\ref{Kramer's solution in 5D}). This assumption guarantees that  the Schwarzschild spacetime is recovered for some particular choices of the parameters. We note that in the cosmological realm, $5D$ Kaluza-Klein  solutions have  been used to generate braneworld  cosmological models (with vanishing bulk cosmological constant) via a relatively simple procedure \cite{Equiv}.

For $b = 0$ and $a = 1$ we return to curvature coordinates and our solutions reduce to some well-known ones in the literature.  For example, the line element (\ref{General temporal Schwarzschild-Kramer-like solution}) reproduces the temporal Schwarzschild spacetime (\ref{T-Schw exterior}) obtained by Casadio, Fabbri and  Mazzacurati \cite{CFM} in search for new black holes on the brane and by Germani and Maartens \cite{Germani} as a possible external metric for an isolated braneworld star. Bronnikov and Kim \cite{Bronnikov 1},  in their example $2$, showed that these spacetimes allow the existence of symmetric traversable wormholes for $r_{0} > 2 m$ ($R_{0} > 2 M$ in our notation). 

For  $b \neq 0$, our solutions display  some interesting physical properties: 
\begin{enumerate}

\item The models with $b < 0$ (Case $1$)  represent traversable wormholes that can have throats located at $R_{0} < 2 M$. What this means is that, as seen from outside, a particle or light falling down reaches the Schwarzschild radius (which no longer defines the horizon) in a finite time. It should be noted that in general relativity, in order to have a throat larger (less) than  the Schwarzschild radius for a given mass at a flat asymptotic, it is necessary to have matter with negative (positive) energy density \cite{B-R Book}.  However, for the braneworld wormholes under consideration here this is not necessarily so.  Indeed, following the reasoning of  \cite{B-R Book} we obtain\footnote{In curvature coordinates $ds^2 = e^{\nu(R)}d t^2 - e^{\lambda(R)} d R^2 - R^2 d\Omega^2$, setting $e^{- \lambda} = 1 - 2 M(R)/R$,  the effective field equation $G_{0}^{0} = 8 \pi T_{0}^{0}$ yields  $M(R) = 4\pi \int{R^2\rho dR} + C$, where $C$ is a constant of integration. Evaluating this expression at the throat $R = R_{0}$ we obtain $C$ in such a way that  $M(R) = M(R_{0}) + 4 \pi\int_{R_{0}}^{R}{R^2\rho dR}$. Now, taking into consideration that $e^{- \lambda(R_{0})} = 1 - 2 M(R_{0})/R_{0} = 0$ and that in the present case $M(\infty) = \gamma M$, which can be checked for all our solutions, we obtain (\ref{relation between the wormhole radius and the effective density}).} 
\begin{equation}
\label{relation between the wormhole radius and the effective density}
R_{0} = 2 \gamma M - 8\pi\int_{R_{0}}^{\infty}{R^2 \rho(R)dR},
\end{equation} 
where $\gamma$ is one of the post-Newtonian parameters in the Eddington-Roberston expansion (\ref{Post-Newtonian parameters}). In general relativity $\gamma = 1$, but in our braneworld solutions it can be less or bigger than $1$ depending  on the choice of various parameters (see, e.g. (\ref{Post-Newtonian parameters for temporal Kramer's solution})). 

\item 
The models with  $b > 0$ (Case $2$) are wormhole spacetimes which have  three extremal spheres with finite area. This is in contrast to standard Lorenzian wormholes, {\it a la} Morris and Thorne, that have only one extremal surface of minimal area which is identified with the throat. 
Here, although the wormholes are symmetrical and twice asymptotically flat the throat is not located at $x = 0$, as in \cite{Bronnikov 1} and \cite{CFM}, but instead is  located at some $x \neq 0$ which explicitly depends on the choice of $\sigma$.  
However, the specific value of the wormhole radius is the same for the whole range $4/3 < \sigma < \bar{\sigma}$; it depends only on the choice of $b > 0$  (see equations (\ref{Rmin for b positive, temporal Schwarzschild-Kramer-like}), (\ref{R + - for the spatial Kramer-like solution modified})).  We note that the extremal spheres have radii 
larger than $2 M$ in the whole range of allowed parameters $b$ and $\sigma$.  These conclusions concerning the Case $2$ are neatly summarized by figure $2$. 

\end{enumerate}

The two-parameter solutions (\ref{General temporal Schwarzschild-Kramer-like solution}), (\ref{spatial Kramer-like solution modified}), (\ref{temporal Kramer-like solution, wormhole}), (\ref{mixed Kramer-like solution, wormhole}) share similar properties, viz., for different values of $b$ (or $a$) and $\sigma$ they can describe black holes, naked singularities and symmetric traversable wormholes of the types discussed in Cases $1$ and $2$. However,  from an experimental point of view they are not equivalent. This follows from the fact that the PPN parameters and the total masses are distinct in each of these solutions. 
Analogous characteristics show the one-parameter solutions (\ref{Kramer-like solution}), (\ref{spatial Kramer-like solution}), (\ref{temporal Kramer-like solution}), (\ref{mixed Kramer-like solution}) which yield the Schwarzschild spacetime for $b = 0$,  naked singularities for $b < 0$ and configurations  that only have one asymptotically flat region for $b > 0$. 

From a formal point of view the symmetric solutions are obtained from those with only one asymptotic region by replacing in the latter 
 $g_{rr} \rightarrow \left(\frac{r - r_{s}}{r - r_{0}}\right) g_{rr}$ and keeping the other metric functions fixed. Here $r_{0} > 2 m $  is an arbitrary parameter, and $r_{s}$ is a constant determined  by the field equations, which in all cases turns to be less than $2 m$. Besides the geometric differences discussed above,  the effective matter is quite different in both cases. In particular, configurations with only one asymptotic region have $T_{0}^{0} < 0 $ everywhere (but their total gravitational mass $M$ is positive), while the symmetric wormholes  have positive effective energy density.

Thus, in this work we have obtained a number of models for wormholes on the brane with interesting physical properties. One can use them to generate new ones by means of iteration.  We note that, although we have restricted our study to the solutions engendered by the Kramer-like metric (\ref{Kramer-like solution}) for which $C = A^{(1 - a - b)/a}$, our discussion can be generalized by considering  $C \propto A^{p}$, where $p$ is some constant parameter (not necessarily $p = (1 - a - b)/a$). With this assumption  we can follow the general approach of Bronnikov and Kim \cite{Bronnikov 1} to study how to choose $A(r)$ in (\ref{field equation for the general spherical metric}) to obtain solutions satisfying wormhole conditions.

The next logical step, to obtain a complete wormhole model within the braneworld paradigm,  is to investigate the extension of our solutions into the bulk.  However, finding an exact solution in $5D$ which is consistent with a particular metric in $4D$ is not an easy task. In spite of this, the existence of such a solution is guaranteed by the Campbell-Magaard's embedding theorems \cite{Seahra}. The coupling of our wormholes solutions to the bulk geometry, though important,  is beyond the scope of the present paper.

\paragraph{Acknowledgments:} I wish  to thank Kirill Bronnikov   for helpful comments and constructive suggestions.


\begin{thebibliography}{99}

\bibitem{WHwiki}{http://en.wikipedia.org/wiki/Wormholes\_in\_fiction}

\bibitem{TTwiki}{http://en.wikipedia.org/wiki/Time\_travel\_in\_science\_fiction} 
\bibitem{geons}{J.A. Wheeler, ``Geons", {\em Phys. Rev.} {\bf 97} (1955) 511.}
\bibitem{Geometrodynamics}{J.A. Wheeler, {\em Geometrodynamics}, Academic Press, New York (1962).}

\bibitem{Morris 1}{M.S. Morris, K.S. Thorne and U. Yurtsewer, ``Wormholes, time machines, and the weak energy condition", {\em Phys. Rev. Lett.} {\bf 61} (1988) 1446.}
\bibitem{Morris 2}{M.S. Morris and K.S. Thorne, ``Wormholes in spacetime and their use for interstellar travel: A tool for teaching general relativity", {\em Am. J. Phys.} {\bf 56} (1988) 395.}

\bibitem{Hayward}{ S.A. Hayward, ``Dynamic wormholes", {\em Int. J. Mod. Phys.} {\bf D 8} (1999) 373 
[arXiv:gr-qc/9805019] }

\bibitem{Novikov}{ N.S. Kardashev, I.D. Novikov and A.A. Shatskiy, ``Astrophysics of wormholes", {\em Int. J. Mod. Phys.} {\bf D 16} (2007) 909 [arXiv:astro-ph/0610441].}

\bibitem{Kuhfittig1}{P.K.F. Kuhfittig, ``Could some black holes have evolved from wormholes?", {\em Schol. Res. Exch.} (2008) 296158  [arXiv:0812.4712].}

\bibitem{Sergey}{S.V. Sushkov and O.B. Zaslavskii, ``Horizon closeness bounds for static black hole mimickers", {\em Phys. Rev.} {\bf D 79} (2009) 067502 [arXiv:0903.1510]. }

\bibitem{Visser 1}{D. Hochberg and  M. Visser, ``The null energy condition in dynamic wormholes", {\em Phys. Rev. Lett.} {\bf 81} (1998) 746 
[arXiv:gr-qc/9802048]. }
\bibitem{Visser 2}{D. Hochberg and  M. Visser, ``General dynamic wormholes and violation of the null energy condition", [arXiv:gr-qc/9901020].}
\bibitem{Visser 3}{ D. Hochberg and  M. Visser, ``Dynamic wormholes, anti-trapped surfaces, and energy conditions", {\em Phys. Rev.} {\bf D 58} (1998) 044021 [arXiv:gr-qc/9802046]. }
\bibitem{Barcelo}{ C. Barcelo and  Matt Visser, ``Traversable wormholes from massless conformally coupled scalar fields", {\em Phys. Lett.} {\bf B466} (1999) 127 [arXiv:gr-qc/9908029]. }
\bibitem{Sushkov 1}{ S.V. Sushkov and S.-W. Kim, ``Wormholes supported by the kink-like configuration of a scalar field", {\em Class. Quant. Grav.} {\bf 19} (2002) 4909   [arXiv:gr-qc/0208069]. }
\bibitem{Kuhfittig2}{ P.K.F. Kuhfittig, ``A single model of traversable wormholes supported by generalized phantom energy or Chaplygin gas
", 	{\em Gen.Rel.Grav.} {\bf 41} (2009) 1485 [arXiv:0904.3566]. }

\bibitem{AP}{K.A. Bronnikov, ``Scalar-tensor theory and scalar charge", {\em Acta. Phys. Pol.} {\bf B 4} (1973) 251.}
\bibitem{Ellis}{H. Ellis, ``Ether flow through a drainhole - a particle model in general relativity", {\em J. Math. Phys.} {\bf 14} (1973) 104.}



\bibitem{Dadhich}{ N. Dadhich, S. Kar, S. Mukherji and  M. Visser, ``R=0 spacetimes and self-dual Lorentzian wormholes", {\em Phys. Rev.} {\bf D 65} (2002) 064004  [arXiv:gr-qc/0109069]. }

\bibitem{Nandy}{K. K. Nandi, B. Bhattacharjee, S. M. K. Alam and J. Evans, ``Brans-Dicke wormholes in the Jordan and Einstein frames", {\em Phys. Rev.} {\bf D 57} (1998) 823.}
\bibitem{Anchordoqui}{ L.A. Anchordoqui, A.G. Grunfeld and D.F. Torres, ``Vacuum static Brans-Dicke wormhole" {\em Grav.Cosmol.} {\bf 4} (1998) 287 [arXiv:gr-qc/9707025].}
\bibitem{Hochberg1}{D.  Hochberg, ``Lorentzian wormholes in higher order gravity theories" {Phys. Lett.} {\bf 251} (1990) 349.}


\bibitem{Hochberg2}{ D. Hochberg, A.  Popov and S. V. Sushkov, ``Self-consistent wormhole solutions of semiclassical gravity", {\em Phys. Rev. Lett.} {\bf 78} (1997) 2050  [arXiv:gr-qc/9701064].}
\bibitem{Remo}{ R. Garattini and  F. S. N. Lobo, ``Self sustained phantom wormholes in semi-classical gravity", {\em Class. Quant. Grav.} {\bf 24} (2007) 2401 [arXiv:gr-qc/0701020].}

\bibitem{BS}{K.A. Bronnikov and A.A. Starobinsky, ``No realistic wormholes from ghost-free scalar-tensor phantom dark energy", {\em JETP Lett.}
 {\bf 85} (2007) 1 [arXiv:gr-qc/0612032].}

\bibitem{Singleton}{ V. Dzhunushaliev and D. Singleton, ``Wormholes and flux tubes in 5D Kaluza-Klein theory",  {\em Phys. Rev.} {\bf D 59} (1999) 064018 [arXiv:gr-qc/9807086].}
\bibitem{Bhawal}{B. Bhawal and S. Kar, ``Lorentzian wormholes in Einstein-Gauss-Bonnet theory", {\em Phys. Rev.} {\bf D 46} (1992) 2464.}
\bibitem{Dotti}{ G. Dotti, J.  Oliva and R.  Troncoso, `` Static wormhole solution for higher-dimensional gravity in vacuum", {\em Phys. Rev.} {\bf D 75} (2007) 024002 [arXiv:hep-th/0607062].}

\bibitem{Bronnikov 1}{ K.A. Bronnikov and Sung-Won Kim, ``Possible wormholes in a braneworld", {\em Phys. Rev.} {\bf  D 67} (2003) 064027  
[arXiv:gr-qc/0212112].}

\bibitem{Lobo}{ F. S. N. Lobo, ``A general class of braneworld wormholes", {\em Phys. Rev.} {\bf D 75} (2007) 064027  
[arXiv:gr-qc/0701133]. }

\bibitem{Randall2}{ L. Randall and R. Sundrum, ``An alternative to compactification", {\em Phys. Rev. Lett.} {\bf 83} (1999) 4690 [arXiv:hep-th/9906064]. }

\bibitem{Shiromizu}{ T. Shiromizu, K. Maeda and  M. Sasaki, ``The Einstein equations on the 3-Braneworld", {\em Phys. Rev.} {\bf D 62} (2000) 024012  [arXiv:gr-qc/9910076]. }

\bibitem{Landau}{L.D. Landau and E.M. Lifshitz, {\em The Classical Theory of Fields}, 4th edn. (Butterworth-Heinemann, 2002).}

\bibitem{Vollick}{ D. N. Vollick, ``Negative Energies on the Brane", {\em Gen. Rel. Grav.} {\bf 34} (2002) 1 [arXiv:hep-th/0004064].}
\bibitem{Bronnikov 2}{ K.A. Bronnikov, H. Dehnen and V.N. Melnikov, ``On a general class of braneworld black holes", {\em Phys. Rev.} {\bf D 68} (2003) 024025  [arXiv:gr-qc/0304068]. }


\bibitem{Visser}{ M. Visser and D.L. Wiltshire, ``On-brane data for braneworld stars", {\em Phys. Rev.} {\bf D 67} (2003) 104004 
[arXiv:hep-th/0212333]. }

\bibitem{Germani}{ C. Germani and R. Maartens, ``Stars in the braneworld", {\em Phys. Rev.} {\bf D 64} (2001) 124010 
[arXiv:hep-th/0107011]. }

\bibitem{CFM}{ R. Casadio, A. Fabbri, L. Mazzacurati, ``New black holes in the brane-world?", {\em Phys. Rev.} {\bf D 65} (2002) 084040 
[arXiv:gr-qc/0111072].}

\bibitem{JPdeL 1}{ J. Ponce de Leon, ``Stellar models with Schwarzschild and non-Schwarzschild vacuum exteriors", {\em Grav. Cosmol.} {\bf 14}   (2008) 65  [arXiv:0711.0998].}
\bibitem{JPdeL 2}{ J. Ponce de Leon, ``Static exteriors for nonstatic braneworld stars", {\em Class. Quant. Grav.} {\bf 25} (2008) 075012 [arXiv:0711.4415]. }

\bibitem{Lemos}{ K.A. Bronnikov and J.P.S. Lemos, ``Cylindrical wormholes", {\em Phys. Rev.} {\bf D 79} (2009) 104019 [arXiv:0902.2360]. }

\bibitem{JPdeL3}{J. Ponce de Leon, ``Kaluza-Klein solitons reexamined", {\em Int. J. Mod. Phys.} {\bf D 17} (2008) 237 [arXiv:gr-qc/0611082]. }

\bibitem{Kramer}{D. Kramer, ``Axialsymmetric stationary solutions of the Projective Field Theory", {\em Acta Phys. Pol.} {\bf B 2} (1970) 807 (In German).}

\bibitem{Weinberg}{S. Weinberg, {\em Gravitation and Cosmology}, John Wiley \& Sons, 1972, page 183.}

\bibitem{Will}{ C.M. Will, ``The confrontation between general relativity and experiment", {\em Living Rev. Rel.} {\bf 4} (2001) 4 
[arXiv:gr-qc/0103036]. }

\bibitem{Equiv}{J. Ponce de Leon, ``Equivalence Between Space-Time-Matter and Brane-World Theories", {\em Mod.Phys.Lett.} {\bf A 16} (2001) 2291  [arXiv:gr-qc/0111011].}
\bibitem{B-R Book}{K.A. Bronnikov and S.G. Rubin, {\em Lectures on Gravitation and Cosmology}, MIFI Press, Moscow, 2008 (in Russian), pages 171-172.}

\bibitem{Seahra}{S. Rippl, C. Romero, R. Tavakol, ``D-dimensional gravity from (D + 1) dimensions", {\em Class.Quant.Grav.} {\bf 12}  (1995)2411, arXiv:gr-qc/9511016; J.E. Lidsey, C. Romero, R. Tavakol, S. Rippl, ``On applications of Campbell's embedding theorem", {\em Class.Quant.Grav.} {\bf 14}  (1997)865, arXiv:gr-qc/9907040; S.S.  Seahra, P.S. Wesson, ``Application of the Campbell-Magaard theorem to higher-dimensional physics", {\em Class. Quant. Grav.} {\bf 20} (2003)1321, arXiv:gr-qc/0302015 ; F. Dahia, C. Romero, ``Dynamically generated embeddings of spacetime",
{\em Class.Quant.Grav.} {\bf 22} (2005)5005, arXiv:gr-qc/0503103.}

\end{thebibliography}
\end{document}